\documentclass[aps,prb,preprint,superscriptaddress,amsmath]{revtex4-1}
\usepackage{graphicx}
\usepackage{epstopdf}
\usepackage{subfig}
\usepackage[capitalize]{cleveref}
\usepackage{tabularx}
\usepackage{setspace}
\usepackage{url}

\begin{document}


\title{A relaxation time model for efficient and accurate 
prediction of lattice thermal conductivity}

\author{Han Xie}
\surname{Xie}
\affiliation{University of Michigan-Shanghai Jiao Tong
 University Joint Institute, Shanghai Jiao Tong University,
 Shanghai 200240, China}
\author{Xiaokun Gu}
\surname{Gu}
\affiliation{Institute of Engineering Thermophysics, 
School of Mechanical Engineering, 
Shanghai Jiao Tong University, Shanghai 200240, China}
\author{Hua Bao}
\surname{Bao}
\email{Corresponding author. E-mail
 address: hua.bao@sjtu.edu.cn (H. Bao).}
\affiliation{University of Michigan-Shanghai Jiao Tong
 University Joint Institute, Shanghai Jiao Tong University,
 Shanghai 200240, China}


\date{\today}

\begin{abstract}
Prediction of lattice thermal conductivity is important to many 
applications and technologies, especially for high-throughput 
materials screening. 
However, the state-of-the-art method based on three-phonon 
scattering process is bound with high computational cost while 
semi-empirical models such as the Slack equation are less accurate.
In this work, we examined the theoretical background of the 
commonly-used computational models for high-throughput thermal 
conductivity prediction and proposed an efficient and accurate 
method based on an approximation for three-phonon scattering strength. 
This quasi-harmonic approximation has comparable computational cost 
with many widely-used thermal conductivity models but had the best 
performance in regard to quantitative accuracy. 
As compared to many models that can only predict lattice thermal 
conductivity values, this model also allows to include 
\textit{Normal} processes and obtain the phonon relaxation time.
\end{abstract}


\maketitle

\section{Introduction}
Lattice thermal conductivity is an important material property 
that plays a key role in many applications and technologies 
\cite{cahill_nanoscale_2002,cahill_nanoscale_2014}. 
For example, heat generation has become a serious issue to further 
improve the performance of semiconductor devices, and thus materials 
with high thermal conductivity are desired for heat dissipation 
\cite{cahill_nanoscale_2002,garimella_thermal_2012,lindsay_first-principles_2013}. 
While in thermoelectric applications, materials with low thermal 
conductivity are more favorable since the thermoelectric performance is 
inversely proportional to thermal conductivity 
\cite{qiu_molecular_2012,yang_thermoelectric_2014,gorai_te_2016}. 
Therefore, finding an efficient and robust method to predict 
lattice thermal conductivity is a desirable goal in itself 
\cite{van_roekeghem_high-throughput_2016,nath_high_2017}. 
With a high-throughput computational framework established, it 
will be much more efficient to search and design new materials 
with tailored thermal conductivities, since we can do materials 
characterization and selection based on theoretical 
understanding before the trial-and-error experimental 
procedures \cite{noauthor_mgi_2011}.
\par
Nevertheless, developing a both accurate and also 
computationally inexpensive method remains a big challenge 
\cite{nath_high_2017,plata_efficient_2017}: accurate methods are 
bound with high computational costs, and fast methods often lack 
the quantitative accuracy. 
As far as we know, the state-of-the-art method for predicting lattice thermal 
conductivity is solving Boltzmann transport equation (BTE) with 
interatomic force constants (IFCs) calculated from first-principles 
calculations. 
The advantage of this method is that it is free of fitting 
parameters and has a good predictive power \cite{lindsay_first_2016}. 
However, extracting anharmonic IFCs from first-principles 
calculations is computationally very expensive. 
Plata \textit{et al.} \cite{plata_efficient_2017} attempted to 
solve such a problem by making effective use of crystal symmetries 
to reduce the number of static first-principles calculations. 
The computational cost is reduced compared with other packages like 
ShengBTE \cite{li_shengbte:_2014} and Phono3py 
\cite{togo_distributions_2015} but it must still be quite large 
because dozens of static first-principles calculations with 
large supercells are still required. 
Compared with ShengBTE, Carrete \textit{et al.} \cite{carrete_almabte_2017} 
developed a similar but more efficient software package named almaBTE 
but the major concern about computational cost of anharmonic IFCs 
remains unaddressed.
\par
Besides the aforementioned method, researchers have tried to 
use some semi-empirical models to predict lattice thermal 
conductivity with less computational cost. 
Among them, the Debye-Gr\"uneisen model \cite{blanco_gibbs:_2004} and 
simplified Debye-Callaway model \cite{miller_capturing_2017} require the least computational resource. 
These two models do not require the computation of harmonic and 
anharmonic IFCs, and therefore have much less computational 
cost. However, the Debye-Gr\"uneisen model 
\cite{blanco_gibbs:_2004} did not show good quantitative 
accuracy when applied to material data sets with different 
structures \cite{toher_high-throughput_2014}. 
It has been implemented in the Automatic-GIBBS-Library (AGL) 
framework in a high-throughput fashion \cite{toher_high-throughput_2014} . 
And we can find that the Pearson correlation coefficient 
between the thermal conductivity calculated with AGL and 
experimental data is high for cubic and rhombohedral 
structures, but significantly lower for anisotropic 
materials and half-Heusler compounds \cite{toher_high-throughput_2014}. 
Miller \textit{et al.} \cite{miller_capturing_2017} tried to 
refine the simplified Debye-Callaway model by introducing four 
fitting parameters and adding the Gr\"uneisen constant. The fitting parameters are dependent on 
the chosen material data set and applying such parameters to 
other materials will be questionable. 
The quasi-harmonic approximation (QHA) is another family of 
methods to predict lattice thermal conductivity, which balance 
between the accuracy and computational cost \cite{nath_high_2017,nath_aflow-qha3p:_2018}. 
Harmonic IFCs are computed to get more accurate Gr\"uneisen 
parameters and Debye temperatures with such methods \cite{nath_high_2017}, 
while the computation of anharmonic IFCs is circumvented. 
Bjerg \textit{et al.} \cite{bjerg_modeling_2014} 
introduced a quasi-harmonic model that uses the 
full dispersion curve computed with harmonic IFCs as the input. 
The model was derived by comparing the Slack equation and the 
Klemens-Callaway model for Debye solids \cite{bjerg_modeling_2014}. 
Nath \textit{et al.} \cite{nath_high_2017} also used the Slack 
equation to calculate lattice thermal conductivity. They tried different 
formulations for the two input parameters and found the best 
combination by comparing their results with experimental values. 
The computational cost of QHA methods are higher than the 
Debye-Gr\"uneisen model or simplified Debye-Callaway model but 
is significantly lower than the full numerical calculation based on BTE. 
Based on the result by Nath \textit{et al.} \cite{nath_high_2017}, 
it can be found that QHA can have better quantitative accuracy 
than the Debye-Gr\"uneisen model and have less fitting 
parameters than the simplified Debye-Callaway model. 
Despite the efforts to refine the semi-empirical models, 
their physical origins are still unclear, e.g. the phase space \cite{lindsay_three-phonon_2008} information is contained in none of them. 
The Slack equation has been widely used in high-throughput 
computation of lattice thermal conductivity but some 
approximations used in its original derivation are unnecessary 
at the present time, including (i) the Debye-like isotropic 
dispersion relation and (ii) a constant function instead of 
Dirac delta function used in its derivation.
\par
In this work, we attempt to find a thermal conductivity 
model that only requires harmonic IFCs as the input, while 
maintains a good prediction accuracy. 
We first reviewed the approximations used in deriving those 
semi-empirical models and tried to identify the necessary 
approximations at the present time. We proposed a model based 
on an approximation developed by Leibfried and Schl\"omann 
\cite{leibfried_heat_1954,leibfried_gittertheorie_1955}, and also 
Klemens \cite{klemens_thermal_1956,klemens_thermal_1958} for 
intrinsic three-phonon scattering strength. This model will use 
the full phonon dispersion data but do not require anharmonic 
IFCs as the input, which can greatly reduce the computational 
cost compared with the full BTE calculation. 
It has comparable computational cost but better accuracy than existing QHA methods. 
This model has been compared with Leibfried and Schl\"omann's 
model, the Slack equation, and Slack's relaxation time model, 
and further discussed.

\section{Theoretical background}
In order to better understand the physical origins of those 
semi-empirical models, we first reviewed the development of the 
theory and the approximations used in their derivation. 
In semiconductors and insulators, phonons are the major heat 
carriers and the lattice thermal conductivity $\kappa_l$ can be 
obtained with the following equation 
\cite{xie_thermal_2014,xie_large_2016}
\begin{equation}
\kappa_l^{\alpha\beta}=\sum_{\lambda}c_\lambda
v^\alpha_\lambda v^\beta_\lambda \tau_\lambda,
\label{eq:k_ab}
\end{equation}
where $\lambda$ denotes different phonon modes that can be distinguished by 
wave vector $\boldsymbol{q}$ and phonon branch $\nu$. $c_\lambda$ is 
the volumetric heat capacity \cite{turney_predicting_2009}. 
$v^\alpha_\lambda$ and $v^\beta_\lambda$ 
are the phonon group velocities in $\alpha$ and $\beta$ directions, 
respectively. $\tau_\lambda$ is the phonon relaxation time.
Among these three phonon properties, $c_\lambda$ and 
$\boldsymbol{v}_\lambda$ are computationally less expensive than 
$\tau_\lambda$ since they only require harmonic IFCs as the input 
while computation of $\tau_\lambda$ needs both 
harmonic and anharmonic IFCs. With harmonic IFCs, we can compute 
phonon frequencies and eigenvectors with harmonic lattice 
dynamics method \cite{turney_predicting_2009}. And then $c_\lambda$ 
and $\boldsymbol{v}_\lambda$ can be calculated with phonon 
frequencies as the input. Under single mode relaxation time 
approximation (SMRTA), the relaxation time can be calculated with \cite{li_shengbte:_2014}
\begin{equation}
\frac{1}{\tau_\lambda^0} = \sum\limits_{\lambda '\lambda ''} {\left( 
{{\Gamma_{\lambda,\lambda',\lambda''} ^{(+)} } + 
\frac{1}{2}{\Gamma_{\lambda,\lambda',\lambda''} ^ {(-)} }} \right)}.
\label{eq:tau_ph0}
\end{equation}
It should be noted that only three-phonon scattering processes are 
considered here and ${\Gamma_{\lambda,\lambda',\lambda''}^{(\pm)}}$ 
are the three-phonon scattering rates for two different types of 
scattering processes \cite{srivastava_physics_1990}. The expressions 
for ${\Gamma_{\lambda,\lambda',\lambda''}^{(\pm)}}$ are \cite{srivastava_physics_1990}
\begin{subequations}
\begin{equation}
\Gamma_{\lambda,\lambda',\lambda''}^{(+)} = \frac{2\pi}{\hbar^2}
\left( {n_{\lambda 
'}^0 - n_{\lambda ''}^0} 
\right)\left|\Phi_{\lambda,\lambda',\lambda''}^{(+)}\right|^2
\delta_{ {\boldsymbol{q}_\lambda} + {\boldsymbol{q}}_{\lambda'} - 
{\boldsymbol{q}}_{\lambda''},{\boldsymbol{G}} }
\delta \left( {{\omega _\lambda } + {\omega _{\lambda '}} - 
{\omega _{\lambda ''}}} \right),
\label{eq:scat_rate1}
\end{equation}
\begin{equation}
\Gamma_{\lambda,\lambda',\lambda''}^{(-)} = \frac{2\pi}{\hbar^2}
{ \left( {1 + 
n_{\lambda '}^0 + n_{\lambda ''}^0} 
\right)\left|\Phi_{\lambda,\lambda',\lambda''}^{(-)}\right|^2
\delta_{ {\boldsymbol{q}_\lambda} - {\boldsymbol{q}}_{\lambda'} - 
{\boldsymbol{q}}_{\lambda''},{\boldsymbol{G}} }
\delta \left( {{\omega _\lambda } - {\omega _{\lambda '}} -
{\omega _{\lambda ''}}} \right)},
\label{eq:scat_rate2}
\end{equation}
\end{subequations}
where $\hbar$ is reduced Plank constant. $n_\lambda^0$ is the 
equilibrium phonon distribution and Bose-Einstein statistics 
should be used. $\Phi_{\lambda,\lambda',\lambda''}^{(\pm)}$ is 
the three-phonon scattering strength.
It should be noted that the first $\delta_{\text{subscript}}$ is 
Knonecker delta function and the second $\delta(~)$ is Dirac delta 
function. $\boldsymbol{G}$ is a reciprocal lattice vector. 
Three-phonon scattering processes with $\boldsymbol{G}=0$ are called
\textit{Normal} processes and those with $\boldsymbol{G}\neq 0$ are 
termed as \textit{Umklapp} processes. 
$\omega_\lambda$, $\omega_{\lambda'}$, and 
$\omega_{\lambda''}$ are the phonon frequencies. 
The three-phonon scattering strength is expressed as
\begin{equation}
\Phi_{\lambda,\lambda',\lambda''}^{(\pm)} = 
\frac{1}{\sqrt{N_0}}
\sum\limits_{b,l'b',l''b''}^{\alpha\beta\gamma}
\Psi_{0b,l'b',l''b''}^{\alpha\beta\gamma}
\sqrt {
\frac{\hbar ^3}{8{m_b}{m_{b'}}{m_{b''}}{
\omega _{\lambda}}
{\omega _{\lambda'}}{\omega_{\lambda''}}}} 
\varepsilon _{b,\lambda}^\alpha \varepsilon 
_{b,\lambda'}^\beta \varepsilon 
_{b,{\lambda''}}^\gamma e^{i \left( \pm \boldsymbol{q}_{\lambda'} 
\cdot {\boldsymbol{R}_{l'}}  - 
{\boldsymbol{q}}_{\lambda''} \cdot 
{{\boldsymbol{R}}_{l''}} \right)} ,
\label{eq:Phi}
\end{equation}
where $N_0$ is the number of $\boldsymbol{q}$-points.
$\Psi_{0b,l'b',l''b''}^{\alpha\beta\gamma}$ is an anharmonic IFC term 
and the subscripts are the atomic indices. For example, $l'b'$ denotes 
the $b'$-th atom in the $l'$-th unit cell. $0$ in the subscript is used to 
denote the center unit cell. $m_b$ is the mass of the $b$-th atom and 
$\varepsilon^\alpha$ is the eigenvector in $\alpha$ 
direction. ${\boldsymbol{R}_{l'}}$ denotes the lattice vector of the 
$l'$-th unit cell. 

\subsection{Leibfried and Schl\"omann's model and the Slack equation}
Equations~(\ref{eq:k_ab})-(\ref{eq:Phi}) were derived many years ago 
but solving them to get a numerical value for lattice thermal 
conductivity was considered to be impossible at that time 
\cite{leibfried_gittertheorie_1955}. 
The major difficulties in solving these equations to get a 
thermal conductivity value lie in two aspects. 
First of all, the three-phonon scattering strength term 
was very complicated. 
In order to get an expression for $\kappa_l$, 
Leibfried \cite{leibfried_gittertheorie_1955} also claimed that the 
eigenvectors had to be analyzed more precisely at that time. 
Secondly, the summation was considered to 
be difficult to carry out \cite{klemens_thermal_1958} and how to deal with Dirac delta function in 
\cref{eq:scat_rate1,eq:scat_rate2} had to be considered 
\cite{leibfried_gittertheorie_1955}. 
\par
Regarding the first issue, Leibfried and Schl\"omann 
\cite{leibfried_heat_1954} derived an approximation for the 
three-phonon scattering strength by generalizing the result of a 
linear chain. 
Later, Klemens \cite{klemens_thermal_1956,klemens_thermal_1958} also 
got a similar formula by generalizing the result for 
long-wavelength phonons to all the phonon modes, where a 
Debye-like dispersion relation and ignorance of phonon branch 
restrictions were also assumed. The quasi-momentum 
conservation rules shown as Kronecker delta functions in 
\cref{eq:scat_rate1,eq:scat_rate2} are the prerequisites to use 
their approximation \cite{klemens_thermal_1958}. 
Srivastava showed a similar equation in his book 
\cite{srivastava_physics_1990}, too. 
All of the three equations are the same except a minor 
difference by a constant factor between each two of them 
\footnote{see Ref. \citenum{klemens_thermal_1956} p. 209 
and Ref. \citenum{srivastava_physics_1990} p. 119}. 
The equation is shown as the following
\begin{equation}
\left| \Phi_{\lambda,\lambda',\lambda''}^{(\pm)} \right| =
B_1 \frac{M\gamma}{\sqrt{N_0}}
\frac{\omega_\lambda \omega_{\lambda'}\omega_{\lambda''}}
{\bar{v}_0}
\sqrt{\frac{\hbar^3}{M^3\omega_\lambda 
\omega_{\lambda'}\omega_{\lambda''}}},
\label{eq:phi}
\end{equation}
where the term 
$\displaystyle\sqrt{\frac{\hbar^3}{M^3\omega_\lambda 
\omega_{\lambda'}\omega_{\lambda''}}}$ is added by us to 
account for the difference between our symbol and 
Klemens's symbol \cite{klemens_thermal_1956}, mainly in the 
representation of creation and annihilation operators. 
$B_1$ is a constant number and
$M$ is the total mass of the atoms in the unit cell 
\cite{roufosse_thermal_1973}. $\gamma$ is the average 
Gr\"uneisen parameter and $\bar{v}_0$ is the 
phonon group velocity in Debye model.
With this estimation, the first issue was solved. 
However, the summation in \cref{eq:tau_ph0} was still considered 
to be difficult in the 1950s, partly due to the second 
issue about the Dirac delta function. 
Leibfried \cite{leibfried_gittertheorie_1955} used a very rough 
approximation to replace the delta function by the inverse of 
Debye frequency $1/\omega_D$. With this estimation the delta 
function smears so broadly that it covers the whole unit 
cell \cite{leibfried_gittertheorie_1955}. 
With these approximations, Leibfried and Schl\"omann 
\cite{leibfried_heat_1954,leibfried_gittertheorie_1955} 
obtained an expression for lattice thermal conductivity shown 
as the following
\begin{equation}
\kappa_l=B_2\left(\frac{k_B\theta_D}{\hbar}\right)^3 
\frac{\bar{M}a_0}{\gamma^2 T},
\label{eq:k_LS}
\end{equation}
where $B_2$ is a constant \cite{leibfried_gittertheorie_1955}.
$k_B$ is Boltzmann constant. 
$\theta_D$ is Debye temperature and is related to Debye 
frequency by $\theta_D=\hbar\omega_D/k_B$. $\bar{M}$ is the 
average atomic mass of the unit cell. 
$a_0$ is the lattice constant and $T$ is temperature.
It should be noted that this equation was very useful and 
convenient in calculating $k_l$ from other known parameters of 
the crystal \cite{slack_thermal_1979} and had been used as a 
standard expression to compare against experimental data \cite{roufosse_thermal_1973}. 
\par
Regarding the value of the constant $B_2$, there exists some 
debates. Julian \cite{julian_theory_1965} claimed that 
Leibfried and Schl\"omann gave a value that is too large by a factor 
of 2 due to a numerical error \cite{julian_theory_1965,klemens_thermal_1986} 
and corrected it to $\displaystyle\frac{3.22}{1.74\times (2\pi)^3}$.
In our calculation with Leibfried and Schl\"omann's model, this 
corrected value was used.
With the help of digital computers, Julian \cite{julian_theory_1965} 
also tried to fit the factor $B_2$ with Gr\"uneisen parameters. 
Based on Julian's fitting parameters, 
Slack \cite{slack_thermal_1979} gave the following equation for 
lattice thermal conductivity
\begin{equation}
\kappa_l=\frac{0.849\times 3\sqrt[3]{4}}
{20\pi^3\left(1-0.514\gamma^{-1}+0.228\gamma^{-2}\right)}
\left(\frac{k_B\theta_D}{\hbar}\right)^3 \frac{\bar{M} 
V^{1/3}}{\gamma^2 T},
\label{eq:k_slack}
\end{equation}
where $V$ is the volume of the unit cell. For 
face-centered cubic structures, $V={a_0}^3/4$. 
This is the Slack equation \cite{slack_thermal_1979} that has 
recently been widely used in the high-throughput computation of 
thermal conductivity by AGL framework \cite{toher_high-throughput_2014}, 
Bjerg \textit{et al.} \cite{bjerg_modeling_2014}, and 
Nath \textit{et al.} \cite{nath_high_2017}. Now the origin of the 
commonly-used Slack equation and the approximations used in its 
derivation are clearly elucidated. It should be noted that the 
expression for $B_2$ in the Slack equation came from a fitting process.
\par
\subsection{Slack's relaxation time model}
Leibfried and Schl\"omann's model and the Slack equation can only 
give the thermal conductivity but physicists may also be interested 
in the information of phonon relaxation times. 
In the 1950s, Klemens, Herring, and Callaway had developed some 
relaxation time models \cite{holland_analysis_1963}. 
Herring \cite{herring_role_1954} developed the formula for 
\textit{Normal} processes. While Klemens \cite{klemens_thermal_1958} 
and Callaway \cite{callaway_model_1959} gave the formulas 
for \textit{Umklapp} processes. At high temperatures (e.g. 
$T\geq 0.1\theta_D$), \textit{Umklapp} processes are the 
dominant scattering mechanism for most of the materials \cite{slack_thermal_1964}. 
And Slack \cite{slack_thermal_1964,glassbrenner_thermal_1964} had 
often used the following model for \textit{Umklapp} processes
\begin{equation}
{\frac{1}{\tau_\lambda^U}=B_U
\frac{{\omega_\lambda}^2 T}{\theta_D} \exp\left(-\frac{\theta_D}{3T}\right)},
\label{eq:tau_model}
\end{equation}
where $B_U$ is a coefficient and Slack 
\cite{slack_thermal_1964,glassbrenner_thermal_1964} obtained 
an expression for it by fitting to the thermal conductivity 
formula given by Leibfried and Schl\"omann. The expression was given as
\begin{equation}
B_U=\frac{\hbar\gamma^2}{\bar{M}{\bar{v}_0}^{2}}.
\label{eq:b_u}
\end{equation}
Bjerg \textit{et al.} \cite{bjerg_modeling_2014} obtained a similar 
expression for $B_U$ by fitting to the Slack equation and their 
formula is different from \cref{eq:b_u} by a factor of about 2.
\subsection{Refinement of the thermal conductivity model}
As we discussed before, the Slack equation has been commonly 
used in high-throughput computation of lattice thermal 
conductivity but not all of the approximations used in its 
derivation are necessary in the present time. 
For example, computation of the full phonon dispersion curve 
was challenging in the 1950s and therefore Debye-like isotropic 
dispersion relation was assumed in its original derivation \cite{klemens_thermal_1958}. 
However, it is not a big challenge now 
and neither is its computational cost very high. 
Nowadays, the full phonon dispersion curve can be used to obtain the 
volumetric heat capacity and group velocity in \cref{eq:k_ab}. 
The summation in \cref{eq:tau_ph0} and the Dirac delta 
function can also be handled easily in numerical simulations. 
Therefore, we propose to still use 
\cref{eq:tau_ph0,eq:scat_rate1,eq:scat_rate2} instead 
of simplified approximations to calculate phonon relaxation times. 
In these equations, the three-phonon scattering strength 
is the computationally most expensive part and we propose to use 
\cref{eq:phi} instead of \cref{eq:Phi} to reduce the 
computational cost. 
\par
With our proposed model, the single mode relaxation time 
\textbf{approximation} used in all of the aforementioned theories also becomes unnecessary. 
Under SMRTA, we need to assume that all of the phonon modes are in 
equilibrium except for just one phonon mode \cite{ziman_electrons_1960}. 
In the 1990s, an iterative method \cite{omini_iterative_1995,omini_beyond_1996,omini_heat_1997} 
was developed as a refinement, which do not need such an assumption. 
The phonon relaxation times can be computed iteratively until 
convergence is reached with relaxation times obtained from 
SMRTA as the initial guess, which is shown below
\begin{equation}
{\tau _\lambda ^f} = \tau_\lambda^0 \left(1 + 
{\Delta _\lambda }\right),
\label{eq:tau_ite}
\end{equation}
where
\begin{equation}
{\Delta _\lambda } = \frac{1}{{{\omega _\lambda }
v_\lambda ^\alpha }}\sum\limits_{\lambda '\lambda ''} {\left[ 
{\left( {v_{\lambda ''}^\alpha {\omega _{\lambda ''}}
\tau_{\lambda ''}^f - v_{\lambda '}^\alpha{\omega_{\lambda'}}
\tau_{\lambda'}^f}\right){\Gamma_{\lambda,\lambda',\lambda''}
^{(+)}} + \frac{1}{2}\left( {v_{\lambda ''}^\alpha {\omega 
_{\lambda ''}}\tau _{\lambda ''}^f + v_{\lambda '}^\alpha 
{\omega _{\lambda '}}\tau _{\lambda '}^f} 
\right){\Gamma_{\lambda,\lambda',\lambda''} ^ {(-)} }} \right]}.
\label{eq:tau_ite2}
\end{equation}
It has to be noted that the superscript $\alpha$ in the equation above 
indicates the direction of thermal conductivity we are interested in.
Iterative method will not introduce tremendous computational 
cost but can incorporate the distinction between 
\textit{Normal} processes and \textit{Umklapp} processes \cite{lindsay_phonon_2014}. 
\par
Finally, we propose to use \cref{eq:tau_ph0,eq:scat_rate1,eq:scat_rate2,eq:phi,eq:tau_ite,eq:tau_ite2} 
to calculate phonon relaxation times iteratively. 
The result from this model is then compared with the original 
result calculated from full iterative method without 
approximations. We also compared our proposed model with \cref{eq:k_LS} 
given by Leibfried and Schl\"omann \cite{leibfried_heat_1954}, 
\cref{eq:k_slack} given by Slack \cite{slack_thermal_1979}, and 
Slack's relaxation time model 
\cite{slack_thermal_1964,glassbrenner_thermal_1964} using 
\cref{eq:k_ab,eq:b_u,eq:tau_model}. The advantages of our proposed model 
are that the computational cost is much lower than the full 
calculation and less approximations are used than those 
semi-empirical models. With our proposed model, more physical 
information is included compared with those commonly-used 
semi-empirical models. 
For example, relaxation times can be extracted from our proposed model. 
The phase space \cite{lindsay_three-phonon_2008} information and \textit{Normal} processes is 
considered in our proposed model while is not 
contained in any of the other models mentioned above.

\section{Simulation details}
In the implementation of our proposed model, we used the 
following equation for the average group velocity 
\cite{chung_role_2004}
\begin{equation}
\frac{1}{\bar{v}_0}=\frac{1}{3}\left(\frac{1}{v_{0,\text{TA1}}}
+\frac{1}{v_{0,\text{TA2}}}+\frac{1}{v_{0,\text{LA}}}\right),
\end{equation}
where $v_{0,\text{TA1}}$, $v_{0,\text{TA2}}$, and 
$v_{0,\text{LA}}$ are the magnitudes of group velocities for 
the three acoustic modes at Brillouin zone $\Gamma$ point, 
including two transverse acoustic (TA) modes and one 
longitudinal acoustic (LA) mode.
Our proposed model was implemented by revising the original ShengBTE 
code. With our proposed model, the scattering 
matrix elements \cite{li_shengbte:_2014} in ShengBTE can be 
derived from the three-phonon scattering strength shown as 
\cref{eq:phi}, which can be expressed as
\begin{equation}
V_{\lambda,\lambda',\lambda''}^{(\pm)} 
= B_1
\frac{\gamma\omega_\lambda \omega_{\lambda'}\omega_{\lambda''}}
{\bar{v}_0}
\sqrt{\frac{8}{M}}.
\label{eq:mat_V}
\end{equation}
It should be noted that the unit conversion factor in the original 
code should also be changed in order to use this equation. 
About the implementation of Dirac delta function, ShengBTE used a 
locally adaptive Gaussian broadening \cite{li_shengbte:_2014}.
\par
Debye temperature was calculated using the expression of Domb and Salter 
\cite{domb_zero_1952,morelli_thermal_2002,nath_high_2017}
\begin{equation}
\theta_D=\sqrt{\frac{5}{3}\frac{\hbar^2}{k_B^2}
\frac{\int_{0}^{\infty}\omega_{\lambda,A}^2 
g(\omega_{\lambda,A})d\omega}
{\int_{0}^{\infty}g(\omega_{\lambda,A})d\omega}},
\label{eq:theta_D}
\end{equation}
where $g(\omega_{\lambda,A})$ is the density of states for the three 
acoustic modes. The integration was replaced by a summation over 
1000 equal intervals from the lowest frequency to the highest frequency. 
The average Gr\"uneisen parameter is calculated with the 
following equation
\begin{equation}
\gamma=\sqrt{\frac{\sum\limits_{\lambda} {\gamma_\lambda}^2 
c_\lambda}{\sum\limits_{\lambda} c_\lambda}}.
\label{eq:gru}
\end{equation}
It should be noted that $\gamma$ is dependent on temperature since 
$c_\lambda$ is related to temperature. 
An average Gr\"uneisen parameter at Debye temperature was used in 
\cref{eq:k_LS,eq:k_slack,eq:b_u} while the average value at temperature $T$ 
was used in \cref{eq:mat_V} or \cref{eq:phi}.
When the Slack equation is used to predict lattice thermal 
conductivity, Nath \textit{et al.} \cite{nath_high_2017} have 
found that the combination of \cref{eq:gru,eq:theta_D} gives 
the best result compared with other expressions for $\gamma$ and $\theta_D$, 
so we adopted these equations to compare with our model.
\par
Three different parameters were used to quantify our model. 
Firstly, the Pearson correlation coefficient $r$ was used to 
measure the linear correlation \cite{nath_high-throughput_2016}. 
Secondly, Spearman's rank correlation coefficient $\rho$ was used to 
assess how well the relationship between two variables can be 
described using a monotonic function \cite{nath_high-throughput_2016}.
Thirdly, we used the average factor difference 
\cite{miller_capturing_2017} (AFD) to quantify the difference 
between the results from different models and the result from 
full calculation, which is given by
\begin{equation}
\text{AFD}=10^x,~x=\frac{1}{n}\sum_{i=1}^{n}\left|\log(t_i)-\log
(p_i)\right|,
\label{eq:AFD}
\end{equation}
where $t_i$ is the true value from full calculation and $p_i$ 
is the predicted value from different models. $n$ is the number 
of samples. The advantage of using AFD is that it gives equal 
weight to all data \cite{miller_capturing_2017}.
\par
Original ShengBTE code package \cite{li_shengbte:_2014} was used to 
compute lattice thermal conductivity from full iterative method. 
A data set of 37 materials was considered and 
the input harmonic and anharmonic IFCs were downloaded from 
almaBTE database 
\footnote{{http://www.almabte.eu/index.php/database}, 
last accessed June 25, 2018.}. 
To have a balanced computational cost, 30$\times$30$\times$30 $q$ mesh was used for materials 
containing two atoms in the primitive unit cell and 
20$\times$20$\times$20 $q$ mesh was used for materials 
containing three 
atoms in the primitive unit cell.
Our proposed model was compared with the full iterative calculation. 
For simplicity, \cref{eq:gru} was used to calculate the average 
Gr\"uneisen parameter with the mode-dependent Gr\"uneisen 
parameters $\gamma_\lambda$ calculated from anharmonic IFCs. 
In real applications of our proposed model, the mode-dependent 
Gr\"uneisen parameters can be obtained from harmonic IFCs, 
which should be the same from those computed from anharmonic 
IFCs \cite{broido_lattice_2005}.
Frequencies and group velocities used in \cref{eq:k_ab} for all the 
phonon modes were calculated within ShengBTE.
$B_1$ was first chosen as $2/\sqrt{3}$ according to 
Klemens \cite{klemens_thermal_1956} and then revised by fitting the 
data to the result from full iterative calculation. 
The acoustic phonon modes were separated by choosing the three 
phonon modes with the lowest frequencies for each 
$q$ point.
No isotope scattering was considered in all of our calculations 
because we are interested in intrinsic lattice thermal 
conductivity in this work.
Our proposed model was then compared with Leibfried and Schl\"omann's 
model, the Slack equation, and Slack's relaxation time model. 
In our implementation of Slack's relaxation time model, 
\cref{eq:k_ab,eq:tau_model,eq:b_u} were used to calculate lattice 
thermal conductivity.
The calculated results from different models were compared using the 
above-mentioned correlation coefficients.

\section{Results and discussion}
\begin{figure}[h!]
\centering
\includegraphics[width=0.8\linewidth]{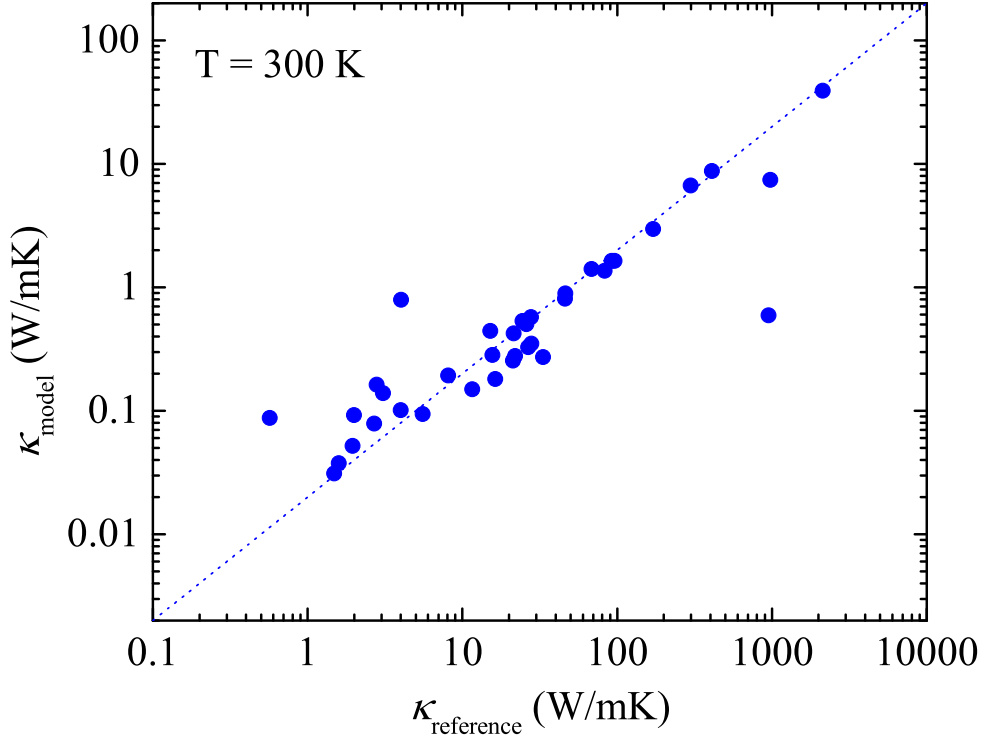}
\caption{Comparison of intrinsic lattice thermal conductivity 
calculated from our proposed model with $B_1=2/\sqrt{3}$ and 
the reference value at 300~K, the dotted blue line is $y=x/50$.}
\label{fig:7result1}
\end{figure}

\begin{figure}[h!]
\centering
\subfloat[]{
\includegraphics[width=0.45\linewidth]{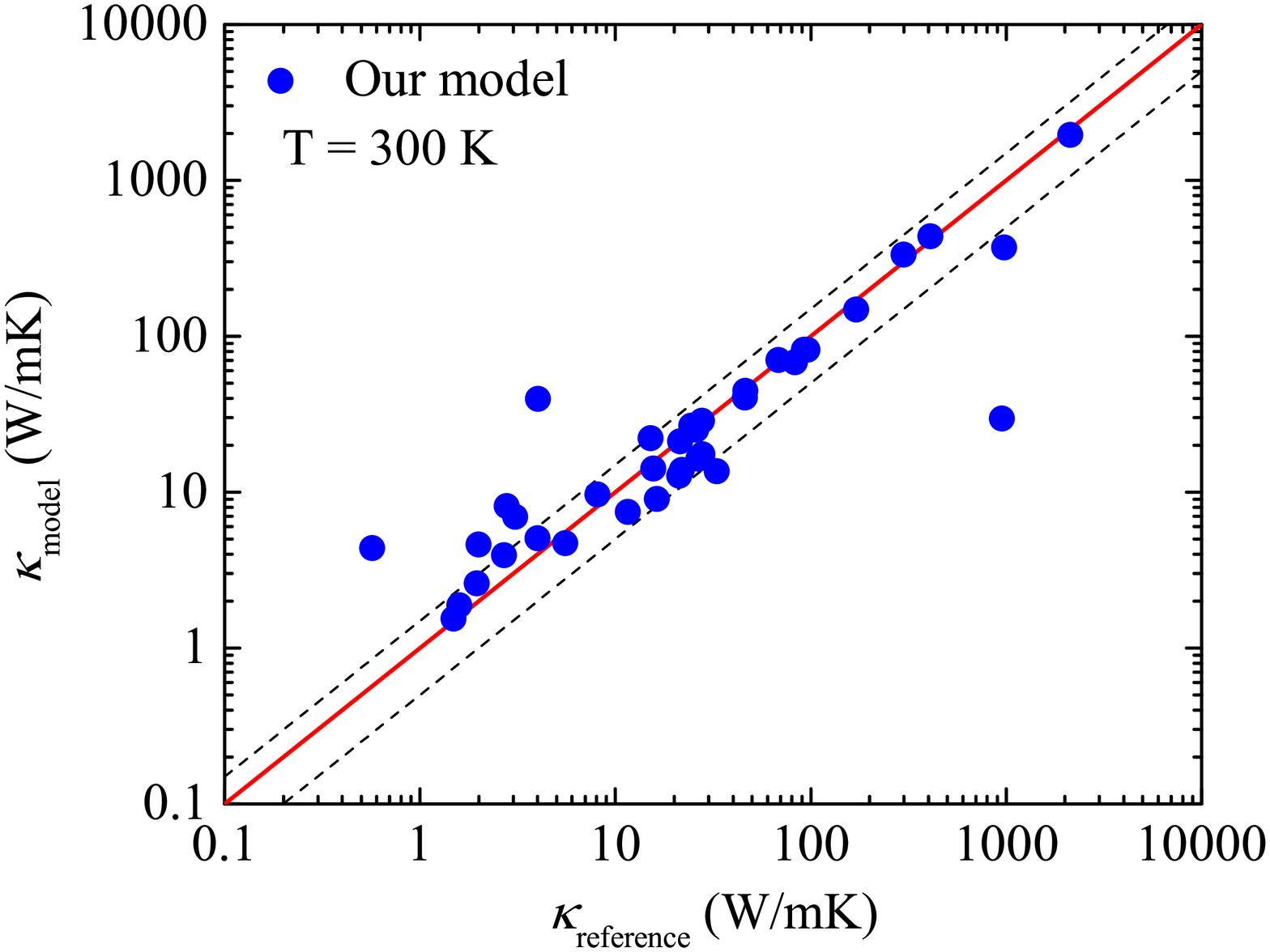}}
\subfloat[]{
\includegraphics[width=0.45\linewidth]{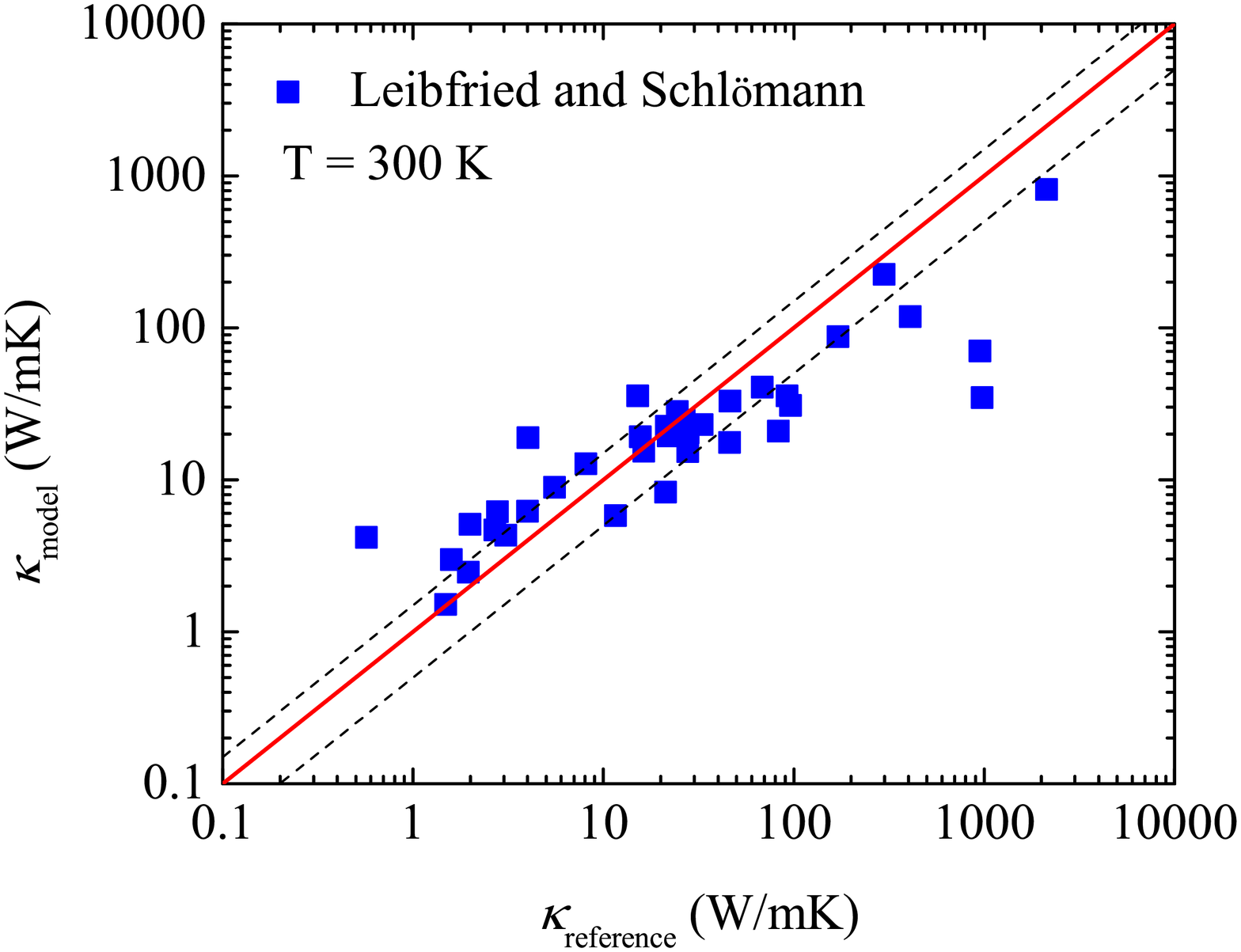}}\\
\subfloat[]{
\includegraphics[width=0.45\linewidth]{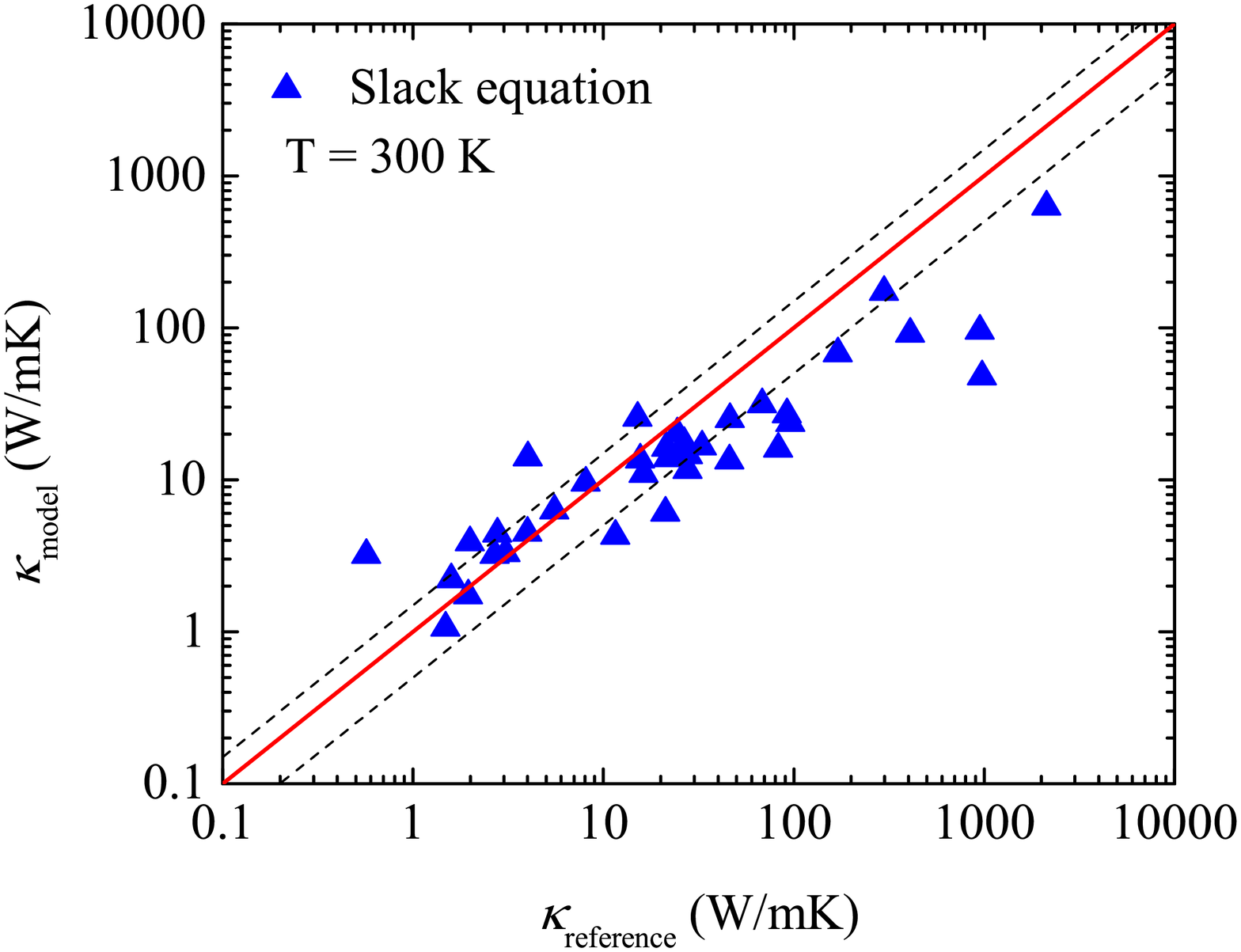}}
\subfloat[]{
\includegraphics[width=0.45\linewidth]{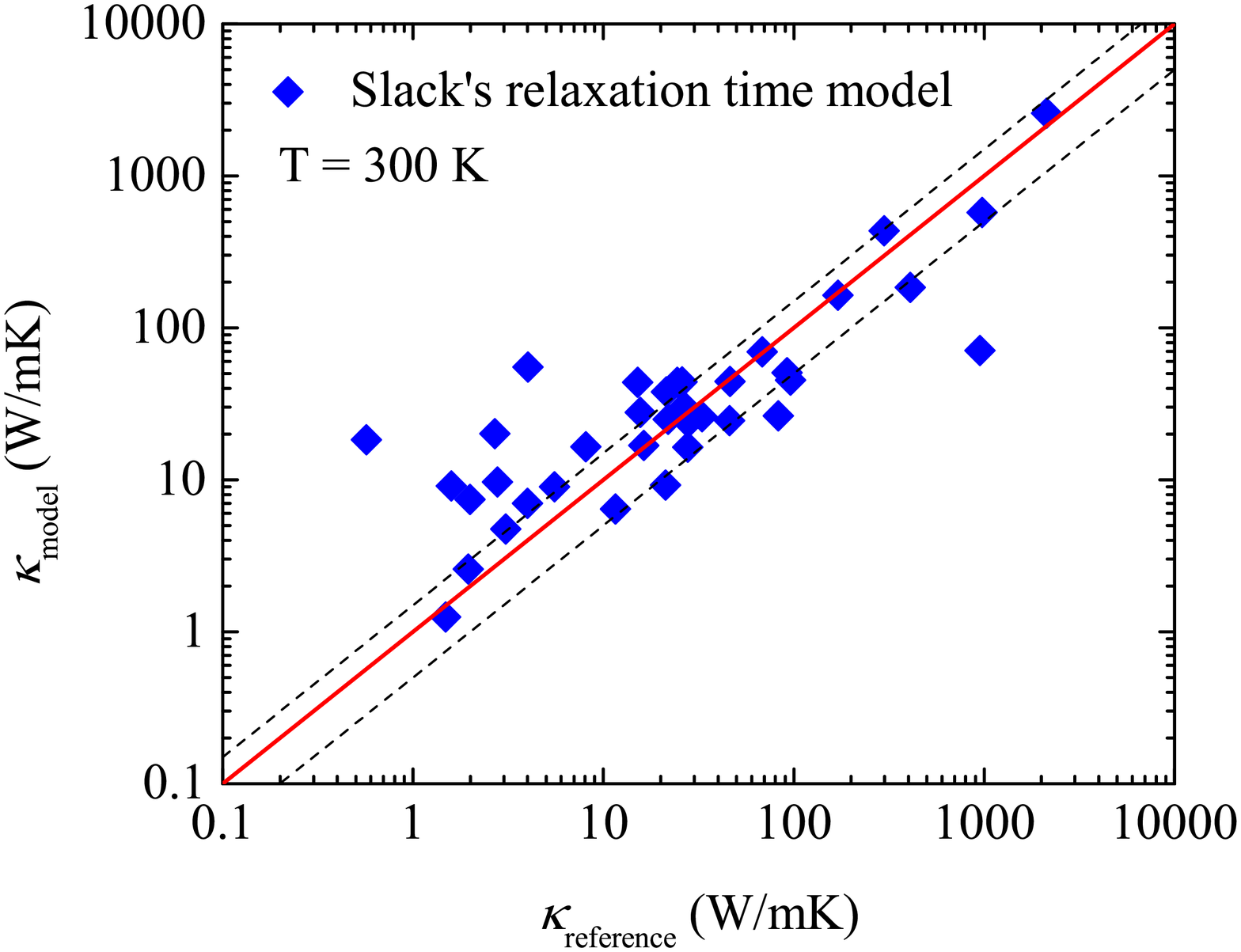}}
\caption{Intrinsic lattice thermal conductivity at 300~K calculated 
from (a) our proposed model with $B_1=2/\sqrt{150}$, (b) Leibfried and 
Schl\"omann's model, (c) the Slack equation, and (d) Slack's 
relaxation time model. 
The solid red line is $y=x$ and dashed black lines represent 
$\pm$50\% error.}
\label{fig:7comparison}
\end{figure}

Intrinsic lattice thermal conductivities at room temperature (300~K) 
calculated from full iterative method were first compared with the results 
from other literatures. 
We found a good agreement between our data and the literature 
\cite{lindsay_ab_2013,carrete_finding_2014}.
The data from full iterative calculation were then used as the reference 
values, which is shown as the $x$ axis in 
\cref{fig:7result1,fig:7comparison}.
The $y$ axis in \cref{fig:7result1} shows the result from our 
proposed model with the initial guess of $B_1=2/\sqrt{3}$. 
The dashed line in \cref{fig:7result1} is the trend line of 
these data and the line equation is $y=x/50$. 
The Pearson correlation coefficient and Spearman correlation 
coefficient between the results from our model and the reference 
values are 0.898 and 0.919, respectively. 
It can be seen that the data calculated from our proposed model 
and those from full iterative method show a very strong correlation. 
\par
However, it was found that there was a quantitative difference in the 
absolute values between $\kappa_\text{reference}$ and our model. 
The reasons for such a difference might be explained as the following: 
Firstly, Klemens \cite{klemens_thermal_1958} claimed that the 
quantitative accuracy of the approximation shown as \cref{eq:phi} might 
not be very reliable. Secondly, the approximations might introduce a 
factor to the expression of $\kappa_l$, which might also be the 
reason why Julian \cite{julian_theory_1965} 
needed to fit the factor $B_2$ with a digital computer. 
As such, it is justified to improve our model by fitting the parameter $B_1$. 
Reducing the constant $B_1$ by a factor of $1/\sqrt{50}$ will make 
the calculated intrinsic lattice thermal conductivity increase to 
50 times its original value. 
Therefore, we suggest to use $B_1=2/\sqrt{150}$ in the future and 
we used this value in our following discussions.
It should be noted that multiplying a constant number to the 
calculated thermal conductivity from our model will not change 
the Pearson and Spearman correlation coefficients.
Even without this adjusted value for $B_1$, the result calculated from our 
model can be used as a very good descriptor for intrinsic lattice 
thermal conductivity. 

\begin{table}
\caption{\label{tab:7correlation}Pearson correlation coefficient $r$, 
Spearman correlation coefficient $\rho$, and average factor difference 
(AFD) between the thermal conductivity calculated from different 
models and the reference value at 300~K.}
\centering
\begin{tabular}{m{0.1\linewidth} m{0.2\linewidth}<{\centering} 
m{0.2\linewidth}<{\centering} 
m{0.2\linewidth}<{\centering} m{0.2\linewidth}<{\centering}}
\hline\hline
 & Our model & Leibfried and Schl\"omann & Slack equation & Slack's relaxation time model \\
\hline
$r$    & 0.898 & 0.865 & 0.896 & 0.906 \\
$\rho$ & 0.919 & 0.897 & 0.908 & 0.815 \\
AFD    & 1.649 & 2.091 & 2.208 & 2.174 \\
\hline\hline
\end{tabular}
\end{table}
\par
Figure \ref{fig:7comparison} shows the comparison of intrinsic lattice 
thermal conductivity from different models at 300~K. We compare our 
model with Leibfried and Schl\"omann's model, the Slack equation, 
and Slack's relaxation time models because they are currently widely used. 
The solid red lines is $y=x$ and is plotted to guide the eye.
The dashed black lines indicate $\pm$50\% error.
The data calculated from these models have been carefully checked 
and the details can be found in the supplementary materials. 
We have fully reproduced the result for silicon calculated by 
Bjerg \textit{et al.} \cite{bjerg_modeling_2014} using the Slack 
equation. Then we moved forward to use our formulas 
for $\gamma$ and $\theta_D$, and further used different models. 
The raw data for these figures can also be found in the supplementary materials. 
From these figures, it can be seen that all these models can predict the 
qualitative trend of $\kappa_l$ reasonably well. As shown in 
\cref{tab:7correlation}, we checked the Pearson and Spearman 
correlation coefficients for different models and found that they are 
consistently larger than 0.86 and 0.81, respectively. 
This also indicates that all of these models can give the qualitative trend. 
By comparing the four models we found that our proposed model 
showed the largest Spearman correlation coefficient and the 
second largest Pearson correlation coefficient. 
We emphasize again here that the Pearson correlation 
coefficient and Spearman coefficient are not related to the 
fitting parameter. Pearson 
correlation coefficient is a measure of the linear relationship 
\cite{nath_high-throughput_2016} but it is sensitive to the 
data distribution \cite{wiki:pearson}. For example, the large 
data points can affect its value significantly. Therefore, in 
order to give an equal weight to all data points, we further 
used AFD to quantify the result. With our fitted 
$B_1=2/\sqrt{150}$, our model shows AFD=1.649, which is 
quite close to unity. By comparison we can find that AFD is 
closer to unity for our model than that for the other models. 
From \cref{fig:7comparison}(a) it can be also seen that most of our data points 
are within the $\pm$50\% error lines. It can be thus concluded that our 
proposed model has the best performance from the comparison. 

\par
\begin{figure}[h!]
\centering
\subfloat[]{
\includegraphics[width=0.45\linewidth]{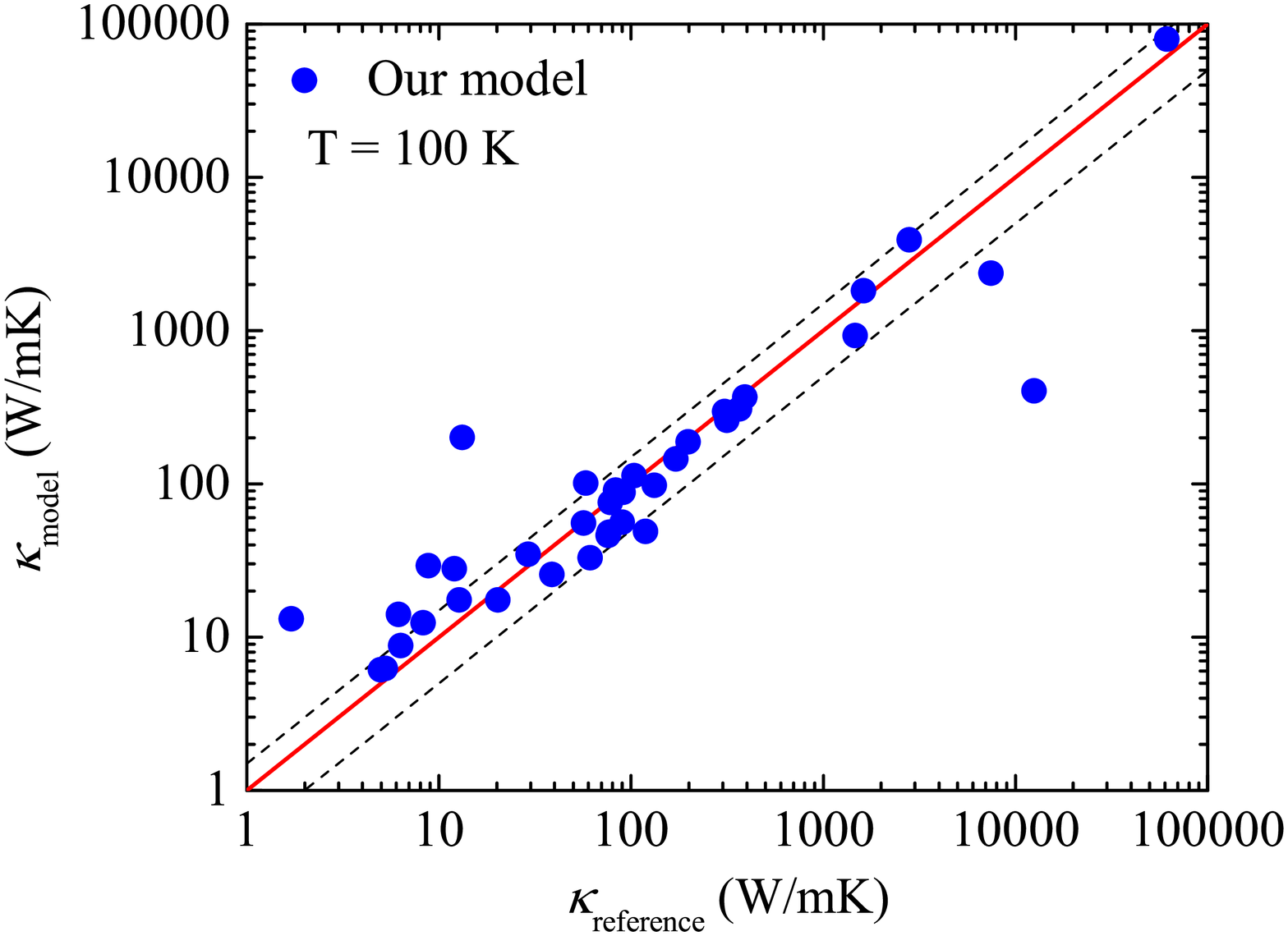}}
\subfloat[]{
\includegraphics[width=0.45\linewidth]{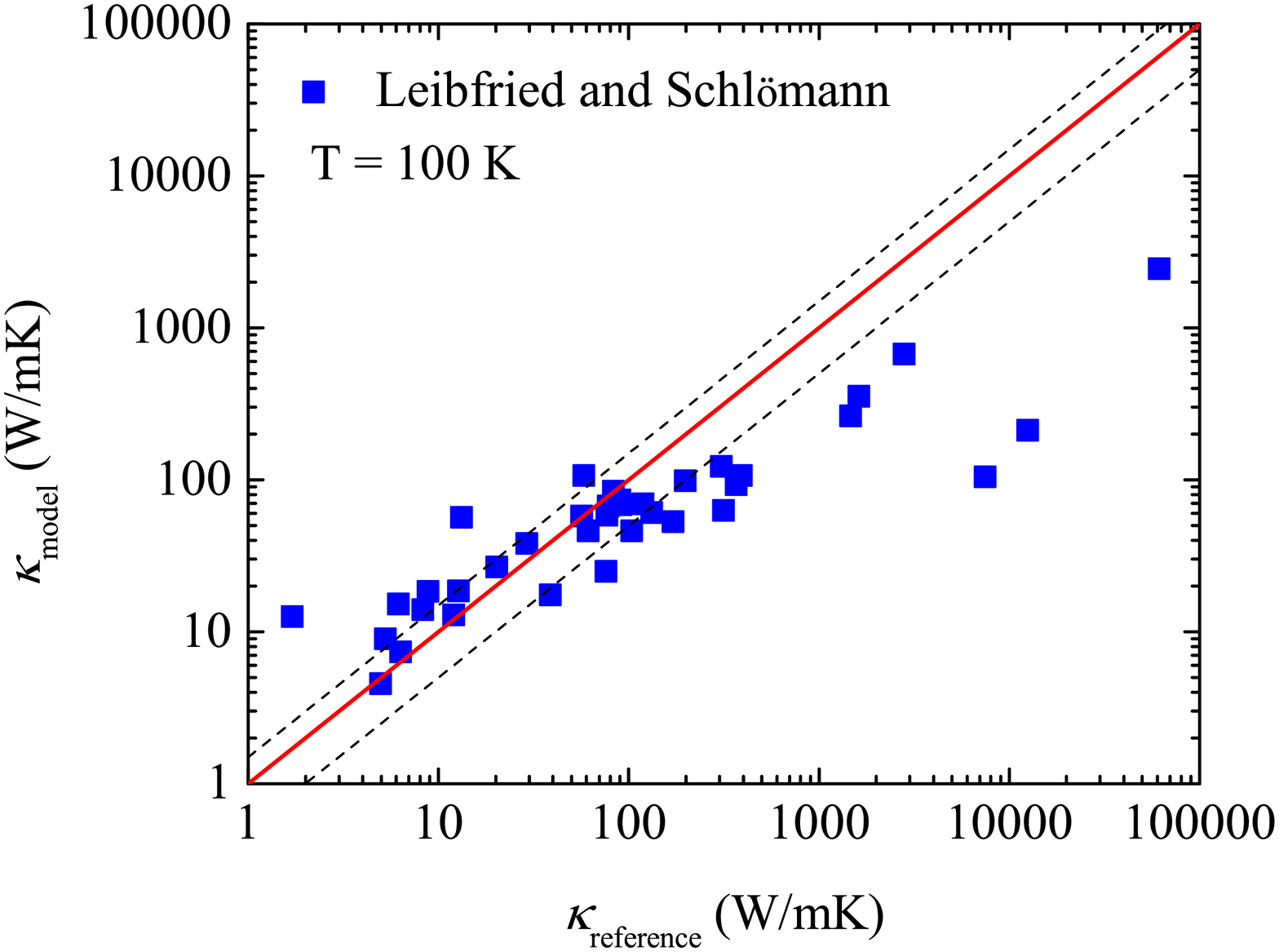}}\\
\subfloat[]{
\includegraphics[width=0.45\linewidth]{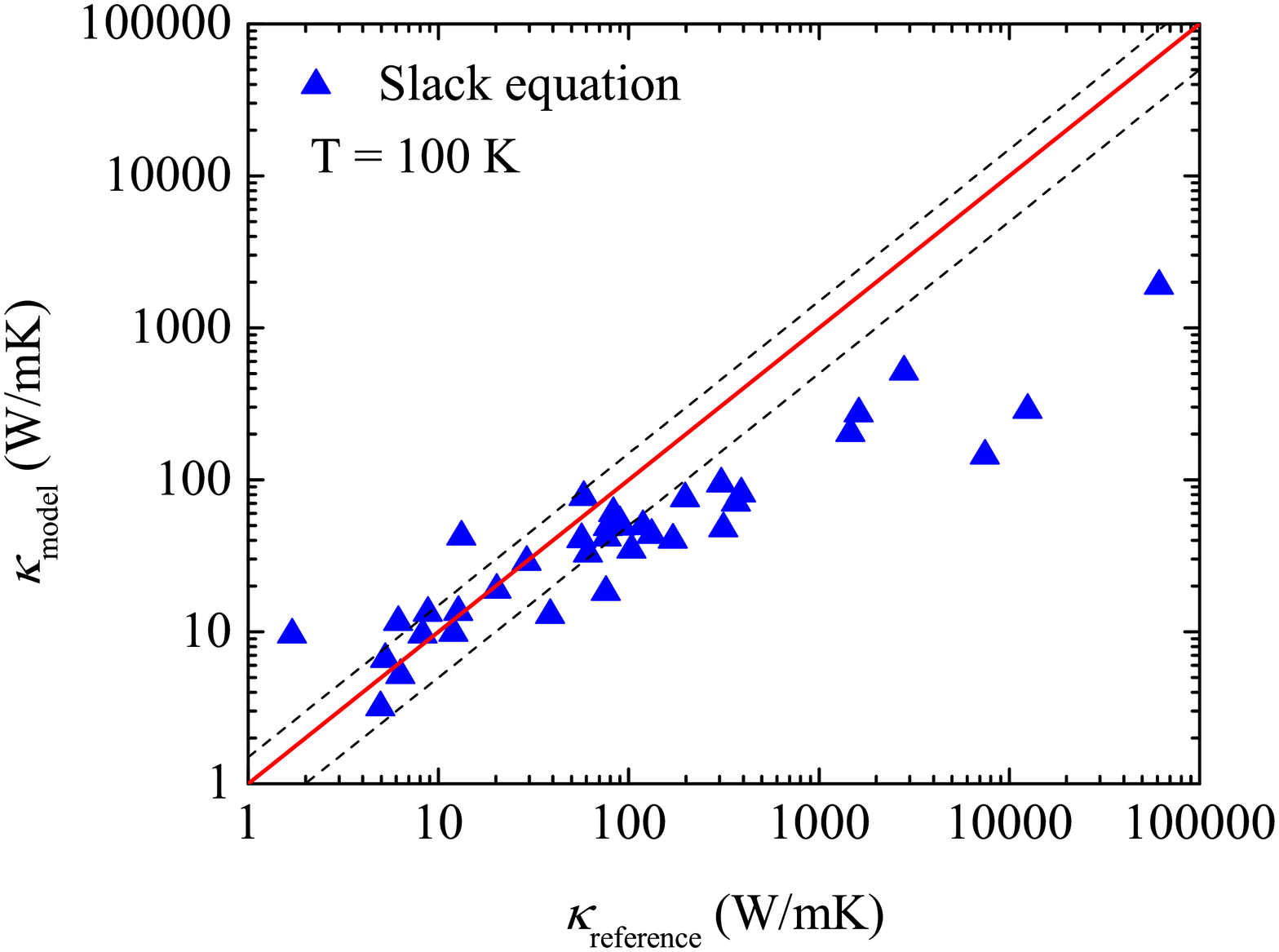}}
\subfloat[]{
\includegraphics[width=0.45\linewidth]{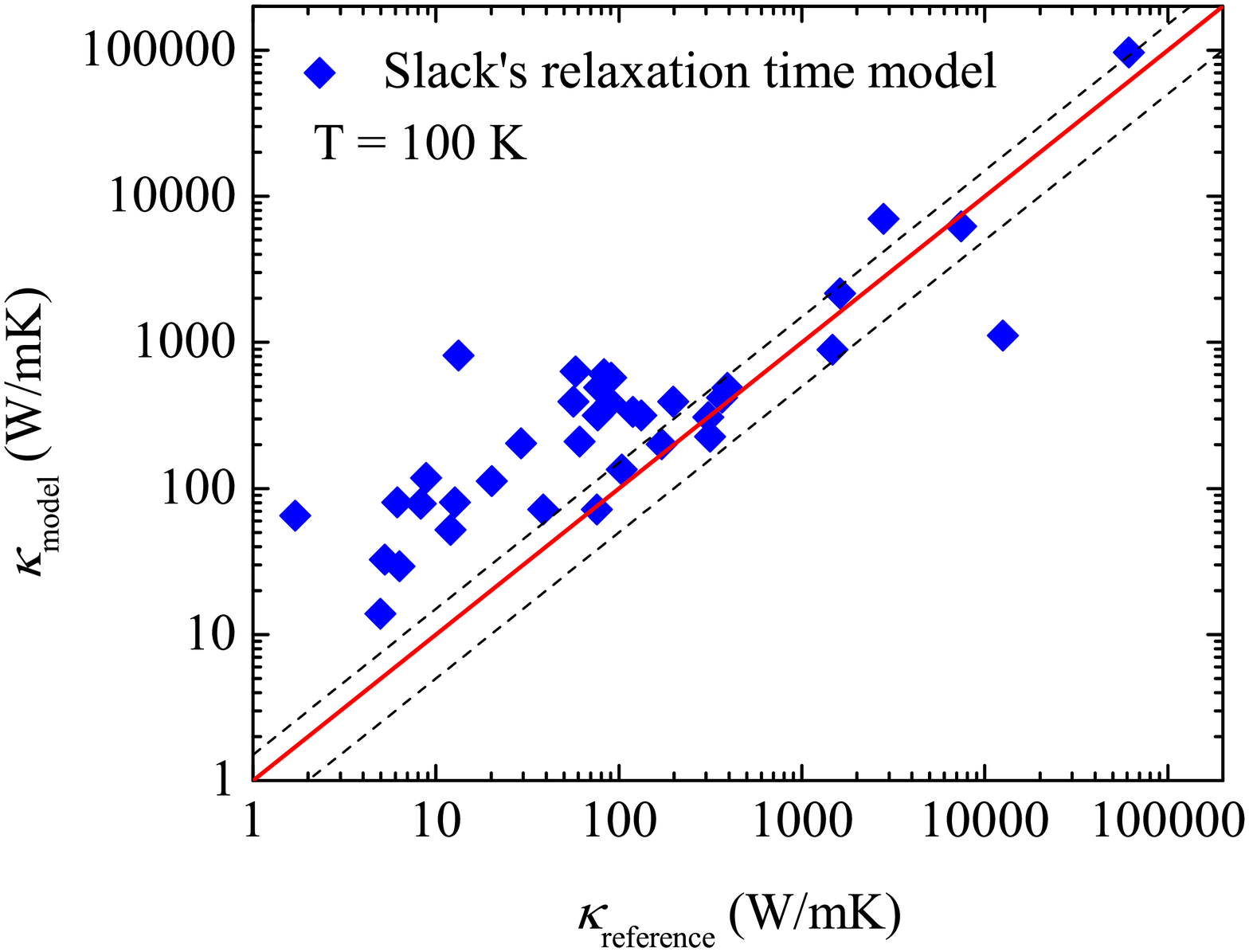}}
\caption{Intrinsic lattice thermal conductivity at 100~K calculated 
from (a) our proposed model with $B_1=2/\sqrt{150}$, (b) Leibfried and 
Schl\"omann's model, (c) the Slack equation, and (d) Slack's 
relaxation time model. 
The solid red line is $y=x$ and dashed black lines represent 
$\pm$50\% error.}
\label{fig:7comparison_100K}
\end{figure}

The reason why our model has the best performance is that we have 
used less approximations than Leibfried and Schl\"omann's model or the
Slack equation, and the full phonon dispersion curve is also taken 
into account. 
The only part where we have used approximations is in calculating the 
relaxation time, to be more specific, in three-phonon scattering 
strength. Besides the better performance, some other advantages of 
our proposed model are: (i) iterative method can be used with our 
model, which would be important when \textit{Normal} processes plays 
a role, (ii) some important phonon information can be obtained from 
our model, e.g. the phonon relaxation times. 
At low temperatures, \textit{Normal} processes will play a more 
important role. To show the importance of \textit{Normal} 
processes, we also showed the result at 100~K in 
\cref{fig:7comparison_100K}. Note that we still kept the same $B_1$ as the 300~K case. 

\begin{table}
\caption{\label{tab:7correlation_100K}Pearson correlation coefficient $r$, 
Spearman correlation coefficient $\rho$, and average factor difference 
(AFD) between the thermal conductivity calculated from different 
models and the reference value at 100~K.}
\centering
\begin{tabular}{m{0.1\linewidth} m{0.2\linewidth}<{\centering} 
m{0.2\linewidth}<{\centering} 
m{0.2\linewidth}<{\centering} m{0.2\linewidth}<{\centering}}
\hline\hline
 & Our model & Leibfried and Schl\"omann & Slack equation & Slack's relaxation time model \\
\hline
$r$    & 0.977 & 0.954 & 0.965 & 0.980 \\
$\rho$ & 0.930 & 0.900 & 0.912 & 0.787 \\
AFD    & 1.730 & 2.686 & 2.953 & 3.841 \\
\hline\hline
\end{tabular}
\end{table}
It can be seen from \cref{fig:7comparison_100K} that our model has the best 
performance. Especially, when compared with Leibfried and 
Schl\"omann's model and the Slack equation, our model has better 
quantitative accuracy for high-thermal-conductivity materials. In 
\cref{tab:7correlation_100K}, we also show the correlation 
coefficients for different models at 100~K. It can be seen that our 
model shows the largest Pearson and Spearman correlation coefficients 
and the smallest AFD. The Pearson correlation coefficient is very 
close to unity because it is sensitive to the largest value in our 
data. However, the other two correlation coefficient, especially AFD, 
do not have such an issue and can be used as a good representation of 
all data points. At 100~K, AFD is 1.730 for our model, which is 
comparable with the result at 300~K. Nevertheless, the other three 
models show much worse AFD at 100~K than that at 300~K. This 
can corroborate the importance of \textit{Normal} processes and the 
good performance of our model at low temperatures.

\begin{figure}[h!]
\centering
\subfloat[Ge at 300~K]{
\includegraphics[width=0.45\linewidth]{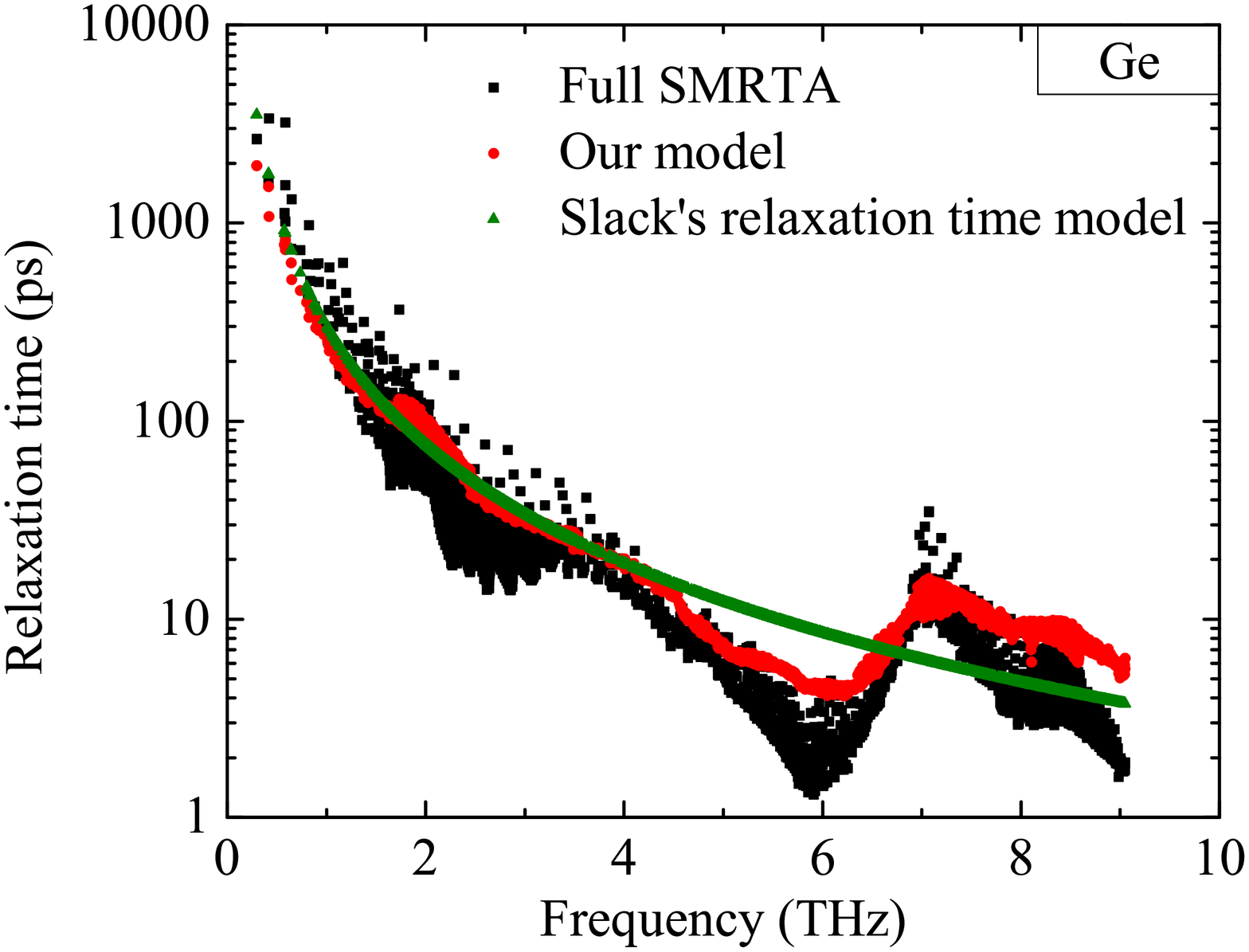}}
\subfloat[Ge at 100~K]{
\includegraphics[width=0.45\linewidth]{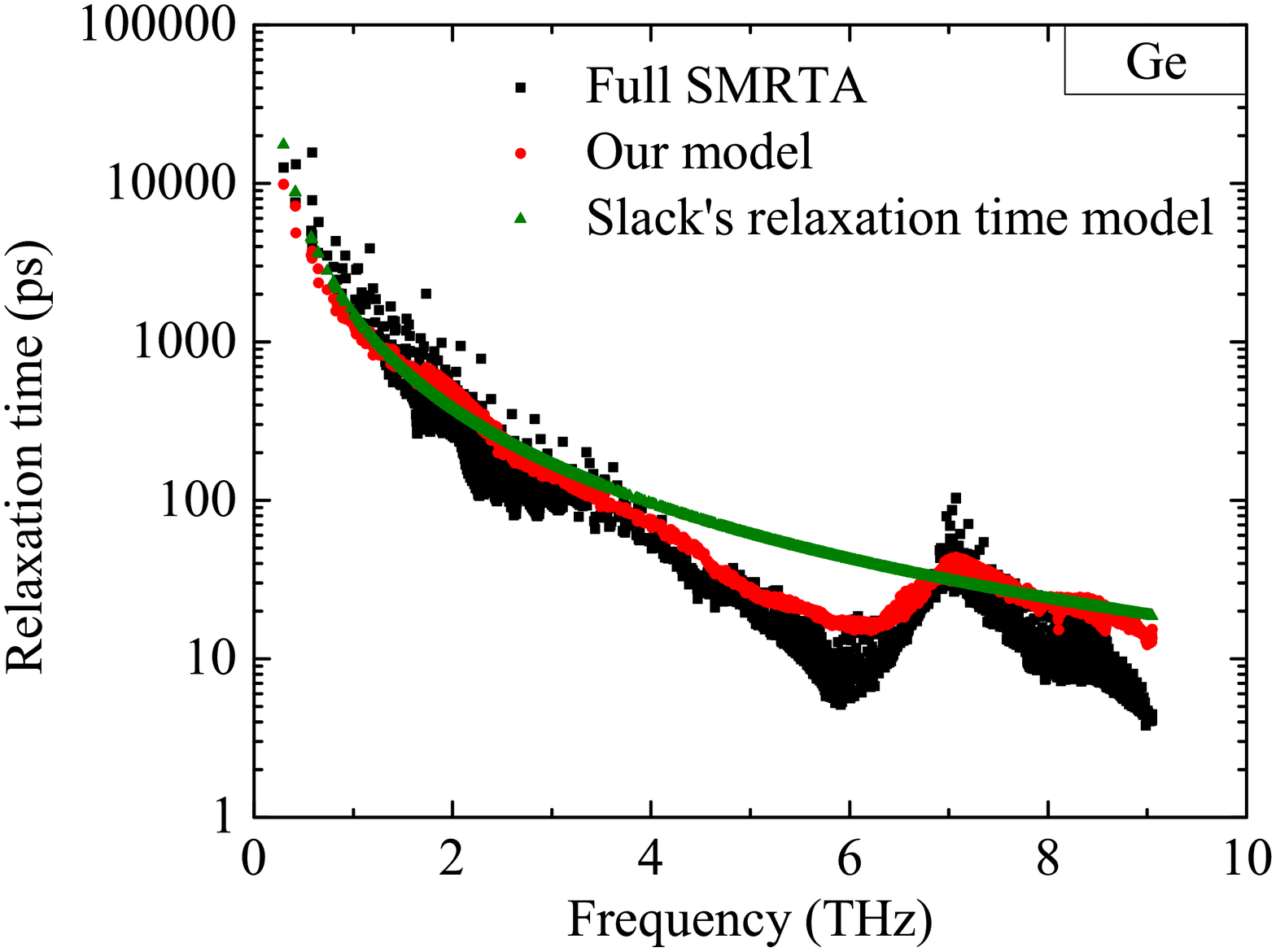}}\\
\subfloat[BN at 300~K]{
\includegraphics[width=0.45\linewidth]{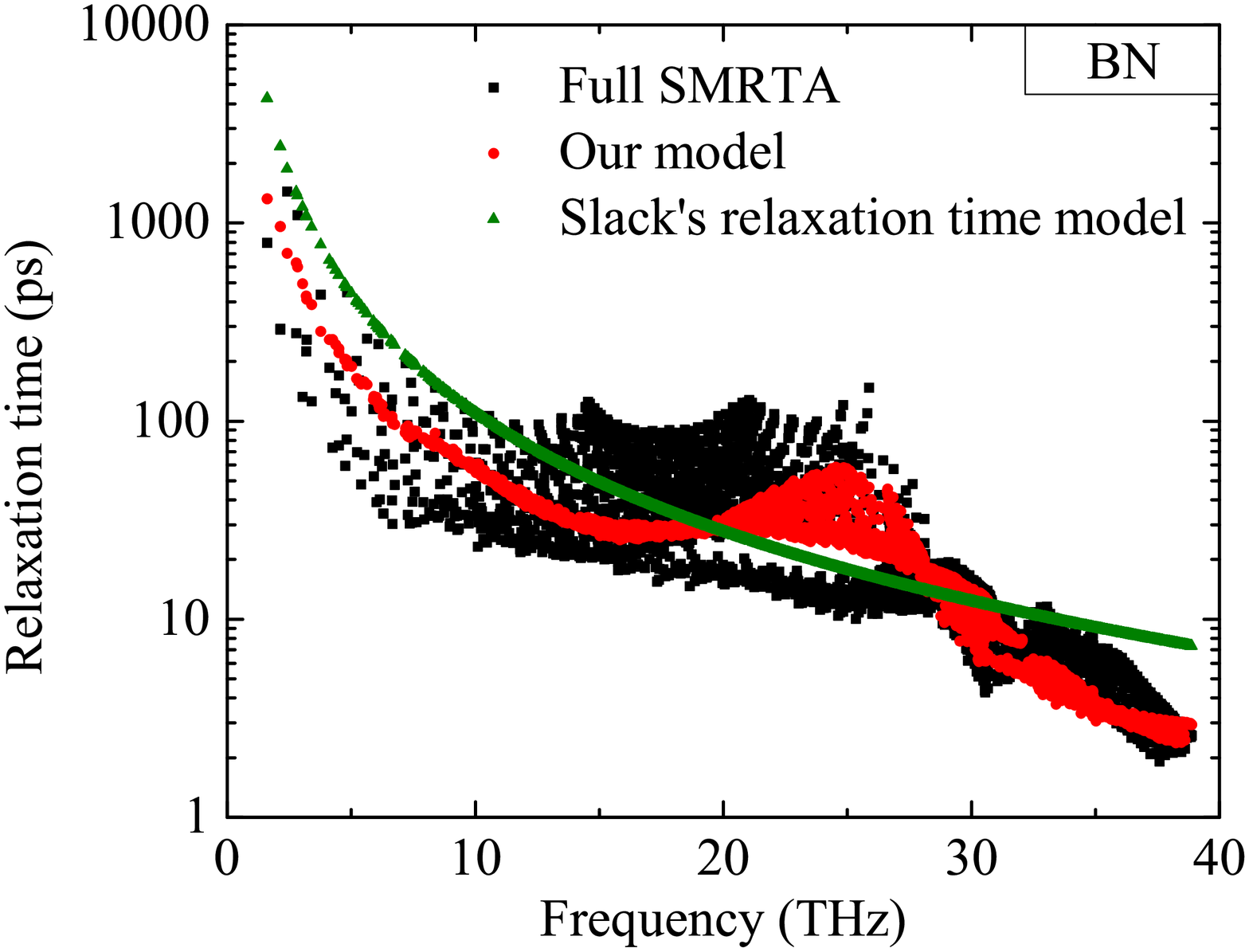}}
\subfloat[BN at 100~K]{
\includegraphics[width=0.43\linewidth]{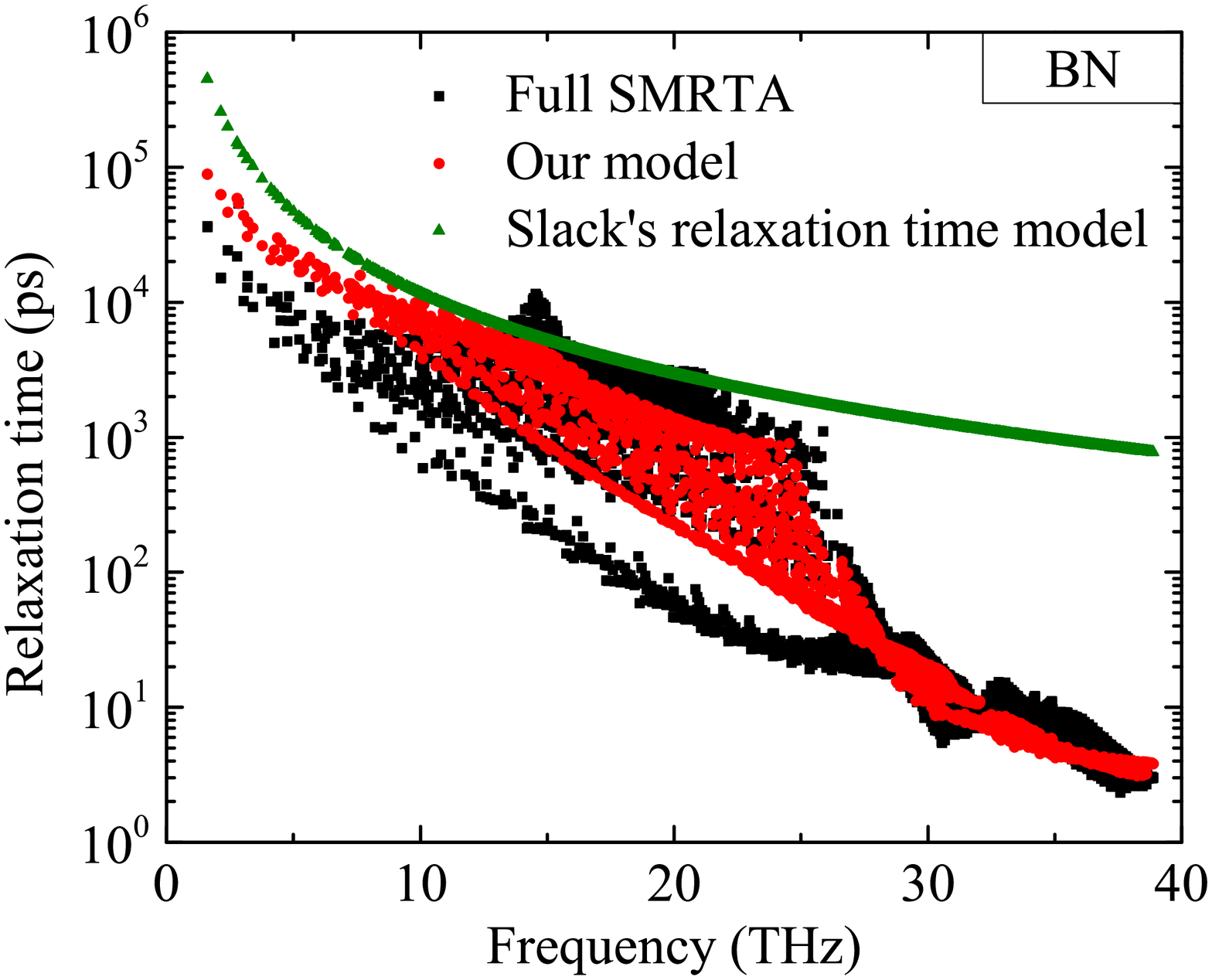}}\\
\subfloat[TeAgLi at 300~K]{
\includegraphics[width=0.45\linewidth]{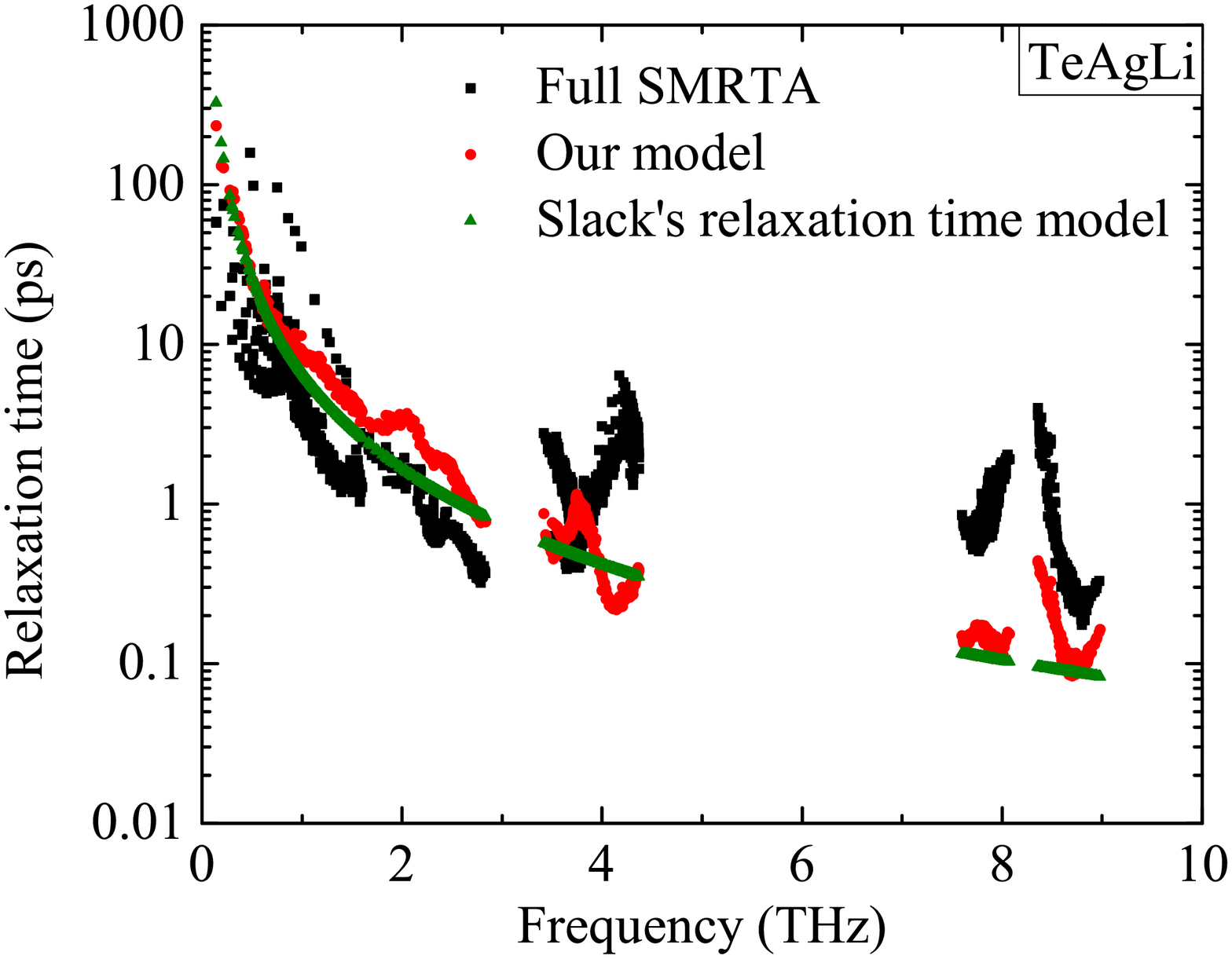}}
\subfloat[TeAgLi at 100~K]{
\includegraphics[width=0.45\linewidth]{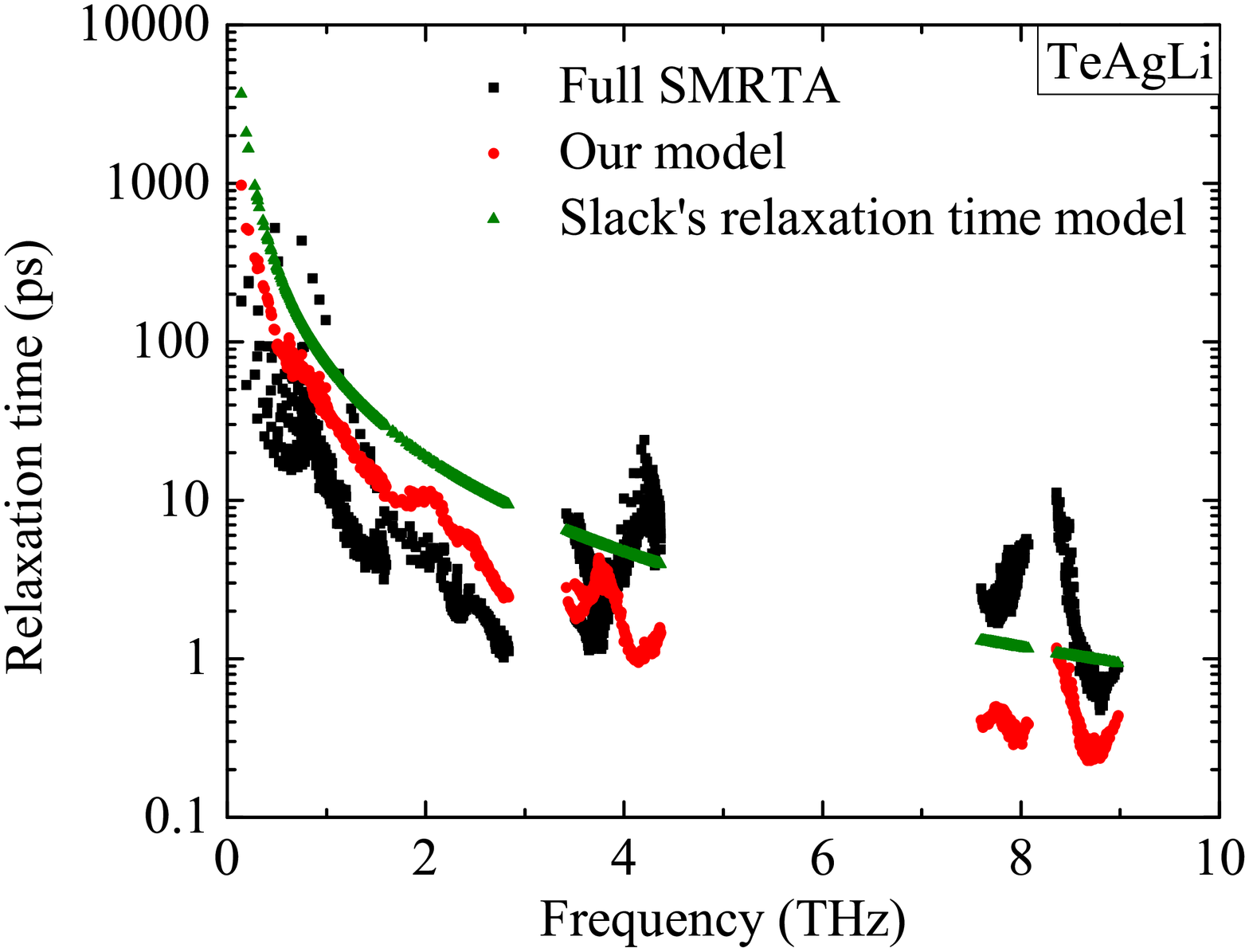}}\\
\caption{Comparison of the relaxation times calculated from full SMRTA method, 
our model, and Slack's relaxation time model.}
\label{fig:tau}
\end{figure}
As we discussed before, another advantage of our model is that some 
important phonon information can be obtained. In \cref{fig:tau}, we 
further compared the relaxation times calculated from full SMRTA 
method, our model, and Slack's relaxation time model. The results for 
three materials were shown in \cref{fig:tau}, where the materials with the highest thermal conductivity 
value (BN), the typical semiconductor material Ge, and a 
half-Heusler compound TeAgLi were chosen. The results for the other 
materials can be found in the supplementary materials. 
The relaxation times from SMRTA method instead of the iterative 
method is used in this figure because relaxation times is not well 
defined in the latter one. 
It can be found that our model can predict the trend of relaxation 
times reasonably well, better than Slack's relaxation time model. 
Slack's relaxation time model can only give a rough trend of the 
relaxation times. 
It should be noted that the $\omega^{-2}$ trend of relaxation 
times in Slack's model may also be questionable \cite{esfarjani_heat_2011}. 
Our model would be physically better in extracting relaxation times 
compared with Slack's model as the scattering phase space 
\cite{lindsay_three-phonon_2008} information is included in our model. 
It should be noted that Lindsay has shown the strong relationship of 
phase space and lattice thermal conductivity \cite{lindsay_first_2016}. 
The relaxation times computed from Slack's model deviates even more 
from the full SMRTA calculation at 100~K than the result at 300~K. 
Our model shows a good agreement with the full SMRTA calculation at 
both 100~K and 300~K. Therefore, our model can better characterize 
the temperature dependence of phonon relaxation times and the lattice thermal conductivity. 
\section{Summary}
In summary, we reviewed the approximations used in deriving the 
Slack equation and identified the necessary approximations at 
the present time. We proposed a model to calculate lattice 
thermal conductivity based on the approximation for 
three-phonon scattering strength, which can be derived from QHA. 
This model is computationally more efficient than the full 
calculation and has comparable computational cost but better 
accuracy than existing QHA methods. 
The full phonon dispersion curve is taken into account in our model.
The results for 37 materials from our proposed model show a strong 
correlation with the calculated thermal conductivities from full 
iterative method. 
We compared our proposed model with other widely-used models, 
including Leibfried and Schl\"omann's model, 
the Slack equation, and Slack's relaxation time model, and 
found that our model has better performance. 
Our model can take \textit{Normal} processes into account and has 
much better performance at low temperatures compared with the other models. 
Another advantage of our model is that some important phonon 
information can be obtained, which will enable us to have better 
understanding about the physics compared with the other 
high-throughput methods like Slack's relaxation time model. The better 
understanding may shed some light on finding 
high-thermal-conductivity or low-thermal-conductivity materials.
Our proposed model finds a balance between accuracy and efficiency 
and can be very useful because of its quantitative predictive power 
and low computational cost compared with the full calculation.

\begin{acknowledgements}
This work was supported by the National Natural Science Foundation of
China (Grant No.~51676121) and the Materials Genome Initiative Center 
of Shanghai Jiao Tong University. Simulations were performed with 
computing resources granted by HPC ($\pi$) from Shanghai Jiao Tong 
University. We thank Jiahao Yan for his help in organizing some data.
\end{acknowledgements}

\section*{Data Availability}
The raw data that support the findings of this study are available 
in the supporting information. The code and scripts for
the method proposed in this study are available from the
corresponding author upon reasonable request.

\section*{Author Contributions}
H.X. carried out calculations with comments and ideas from X.G. and H.B. All 
authors analyzed the data. H.B. supervised the research. H.X. prepared the draft manuscript. X.G. and H.B. commented on, discussed and edited the manuscript.

\section*{Competing Interests}
The authors declare no competing financial or non-financial interests.

\end{document}